\documentclass[10pt, conference, letter]{IEEEtran}
\IEEEoverridecommandlockouts
\usepackage{graphicx} 
\usepackage{color}
\usepackage{amsmath}
\usepackage{comment}
\usepackage{multirow}
\usepackage[table,xcdraw]{xcolor}
\usepackage{hhline}
\usepackage{flushend}

\title{Trading BRAM for LUTs: Hybrid Bit-Level Weight Storage for FPGA-Based Spiking Neural Networks}

\author{
\IEEEauthorblockN{Dimitrios Sekertzis and Giorgos Dimitrakopoulos }
\IEEEauthorblockA{ 
Integrated Circuits Lab, Electrical and Computer Engineering, Democritus University of Thrace (DUTH), Greece}}

\begin{document}

\maketitle

\begin{abstract}
Spiking Neural Networks (SNNs) offer an attractive paradigm for energy-efficient AI inference, while their deployment on reconfigurable hardware enables application-specific acceleration without sacrificing model reprogrammability. However, realizing this potential on FPGAs is often severely constrained by the immense on-chip memory needed to store synaptic weights. Existing compression techniques reduce this storage requirement at the cost of weight precision or accuracy. This work presents a hybrid bit-level weight-storage scheme that trades BRAM capacity for FPGA logic resources while preserving the original weights unmodified. Selected weight bits are implemented using Look-Up Tables (LUTs), while the remaining bits are stored in BRAM and jointly reconstructed during inference. We integrate this approach into a spatial dataflow SNN architecture and introduce a practical bit-mapping heuristic that efficiently partitions weight bits between BRAM and LUT-based logic without modifying the weights. Experimental evaluation demonstrates more than 26\% reduction in BRAM utilization with a modest increase in LUT and register usage, resulting in significant overall power savings.
\end{abstract}

\section{Introduction}
The increasing energy demands of conventional Machine Learning have renewed interest in Spiking Neural Networks (SNNs), whose sparse, event-driven computation offers an attractive path to highly energy-efficient AI inference. Realizing this potential while retaining the flexibility to deploy different SNN models on a single chip presents a critical architectural choice. Programmable neuromorphic ASICs, such as Loihi2~\cite{loihi2} and SpiNNaker2~\cite{spinnaker2}, provide general-purpose flexibility, but their complex routing networks and software-like programmability introduce substantial hardware overhead that can limit energy efficiency. FPGAs offer an attractive balance between efficiency and programmability by leveraging a reconfigurable fabric without embedding general-purpose neuromorphic infrastructure into the accelerator itself~\cite{firefly, s2n2, fpga-snn}. This enables developers to synthesize fully application-specific SNN architectures tailored to the exact topology of a target network for maximum efficiency, 
while retaining the ability to deploy different models via FPGA bitstream reconfiguration~\cite{spiker, modnef}.

Despite this ideal blend of structural efficiency and hardware flexibility, scaling FPGA-based SNN accelerators in practice is frequently bottlenecked by the finite availability of on-chip memory~\cite{modnef, layerspec}. In addition to tracking neuronal states and intermediate data, the accelerator must accommodate an immense number of synaptic weights. Keeping this complete network topology on-chip is imperative, as fetching weights from external memory drastically increases power and inference latency thus negating the core benefits of the SNN~\cite{memory_review}. Consequently, minimizing the on-chip memory footprint of synaptic weights has emerged as a critical requirement for scalable and efficient FPGA-based SNN inference.

Existing approaches typically address this limitation by reducing the amount of information required to represent the weights. Techniques such as pruning~\cite{pruning}, quantization~\cite{flowq}, sparse weights~\cite{fpspike}, and vector quantization~\cite{vq4snn} originally developed extensively for deep neural networks, have also been adopted for SNNs. Pruning eliminates weights considered less important, quantization represents weights using fewer bits, weight sparsity encodes only non-zero weights with appropriate metadata, and vector quantization maps groups of weights to compact indices in a shared codebook.
While these techniques reduce storage requirements, they trade memory savings for inference accuracy. 

In this work, we propose a different approach to reduce the number of weight bits stored in FPGA BRAMs without modifying the weights or sacrificing precision. The key idea is to exploit the heterogeneous nature of FPGA resources by distributing the bits of each weight between memory and programmable logic: \emph{selected bits are generated online using LUTs, while the remaining bits are stored in BRAM}. The complete weight is reconstructed from these two sources during inference, effectively trading BRAM capacity for LUT resources while preserving the original weight representation. While embedding parameters and operations into LUTs is established for conventional logic-based neural networks~\cite{pedram, neuralut}, these stateless combinational techniques cannot support the stateful membrane potentials and temporal feedback inherent to SNNs~\cite{snn-vs-ann}. Consequently, using LUTs strictly for online weight generation is a novel, SNN-specific optimization.

We integrate the proposed hybrid weight-storage scheme into a spatial dataflow SNN architecture~\cite{s2n2,modnef, spiker_plus}, where network layers and neurons are physically mapped onto the FPGA. At design time, we use a practical heuristic to efficiently partition the weight bits between BRAM and LUT-based logic independently per network layer. 
The main contributions of this work are summarized as follows:
\begin{itemize}
\item
A hybrid weight-storage scheme for FPGA-based SNNs that distributes synaptic-weight bits between BRAM and LUT-based online weight generation logic, while preserving the original weight precision. A practical heuristic determines the bit allocation between the two resources.
\item
A spatial data flow SNN architecture supporting the proposed split-storage paradigm, where actual weights are reconstructed on the fly from BRAM-stored and logic-generated bits.
\item
An FPGA SNN implementation and evaluation against state-of-the-art accelerators across representative SNN models, demonstrating substantial BRAM savings of over 26\% in all cases at the cost of a modest increase in LUT and register utilization. While power improves across all examined cases, when focusing on comparisons that refer to SNNs implemented on the same FPGA device power is reduced by 45\% on average.

\end{itemize}

\section{SNN Fundamentals and Spatial-Dataflow FPGA Architecture}
SNNs process information temporally by communicating through discrete, asynchronous binary pulses known as spikes. To process continuous real-world data, input values must first be mapped into spike trains over a defined number of time steps. This input encoding is typically achieved using rate coding (where spike frequency is proportional to the input intensity), temporal coding (where precise spike timing carries information), or direct encoding (where analog inputs are applied directly to the first layer's membrane potentials). Once encoded, these spike trains propagate through the network's topology, driving the internal dynamics of the neurons.

\begin{figure}[t]
\centering
\includegraphics[width=0.75\columnwidth]{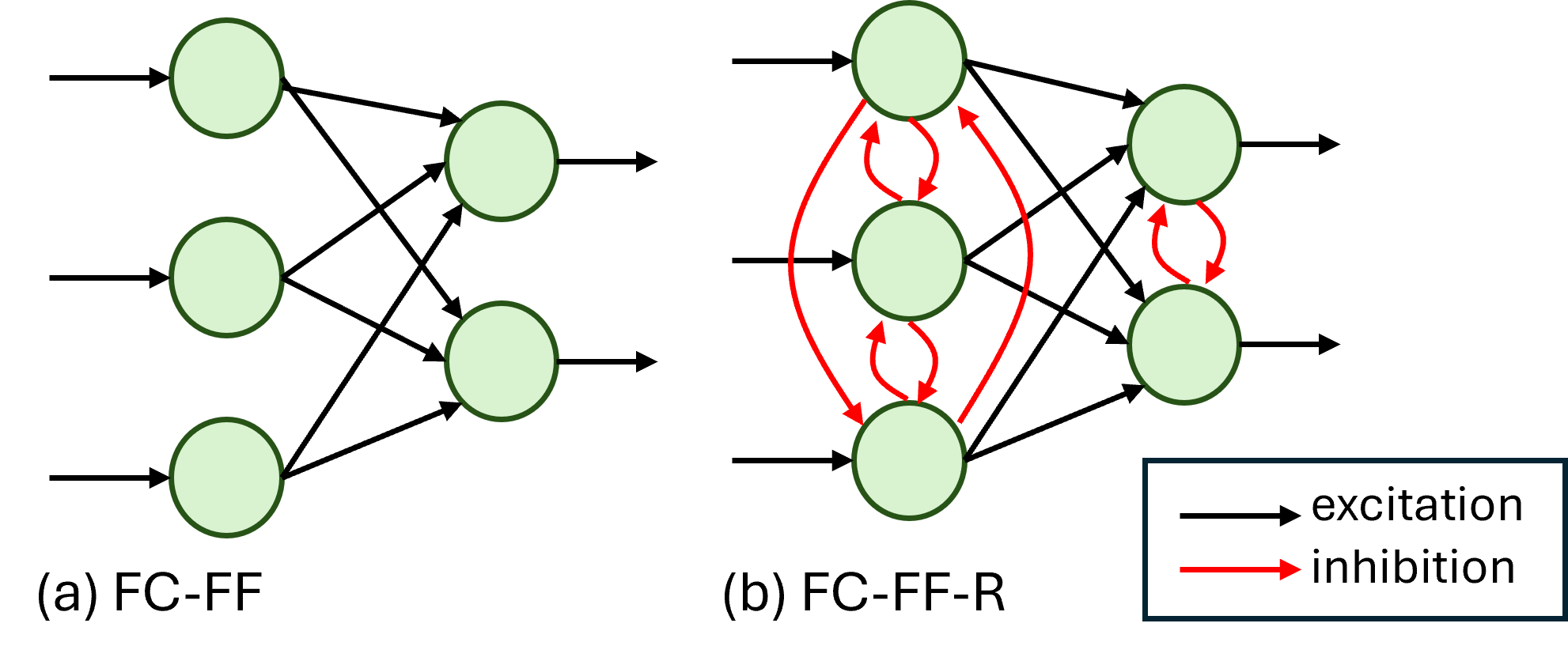}
\caption{SNN topologies: (a) fully connected (FC) feed-forward (FF) network; (b) fully connected (FC) feed-forward (FF) network with recurrent (R) lateral inhibition.}
\label{f:snn}
\end{figure}

\subsection{Leaky Integrate-and-Fire Neuron Model}
The Leaky Integrate and Fire (LIF) model is widely used in hardware implementations due to its balance of biological plausibility and computational efficiency, abstracting away the prohibitive complexity of high-fidelity models like Hodgkin-Huxley and Izhikevich~\cite{survey, fpga_survey}. A LIF neuron integrates incoming spikes by accumulating their corresponding synaptic weights into its membrane potential, while continuously leaking a portion of that potential over time toward a resting state.

In LIF, the discrete-time update of the membrane potential $V_j[t]$ for neuron $j$ at time step $t$ is formulated as:
\begin{equation}
V_j[t] = V_j[t-1] + \sum_{i} w_{ji} s_i[t] - \lambda V_j[t-1]
\end{equation}
Here, $\lambda$ represents the leak factor, and the second term is the inner product of the synaptic weights $w_{ji}$ and the incoming binary spikes $s_i[t]$ from pre-synaptic neurons. To minimize hardware complexity, the leak factor $\lambda$ can be approximated using negative powers of two (e.g., $\lambda = 2^{-L}$), with minimum accuracy loss, as shown in~\cite{spiker_plus} allowing the decay to be implemented efficiently with simple bitwise right shifts rather than costly multipliers.

After the state update, the membrane potential is compared against a firing threshold $V_{\mathrm{th}}$. An output spike $s_j[t]$ is generated whenever the potential reaches or exceeds this threshold:
\begin{equation}
s_j[t] = 
\begin{cases} 
1, & V_j[t] \geq V_{\mathrm{th}},\\ 
0, & V_j[t] < V_{\mathrm{th}}. 
\end{cases}
\end{equation}
Following the generation of a spike, the membrane potential must be reset to prevent the neuron from continuously firing. In a subtractive (soft) reset, the threshold value is simply subtracted from the current membrane potential, retaining any residual charge. Alternatively, a hard reset forces the membrane potential to a predefined resting state. While the choice of reset mechanism affects the temporal dynamics of the spike train, a hard reset to a fixed trainable reset parameter is adopted across all models in this work.

\begin{figure}[t]
\centering
\includegraphics[width=\columnwidth]{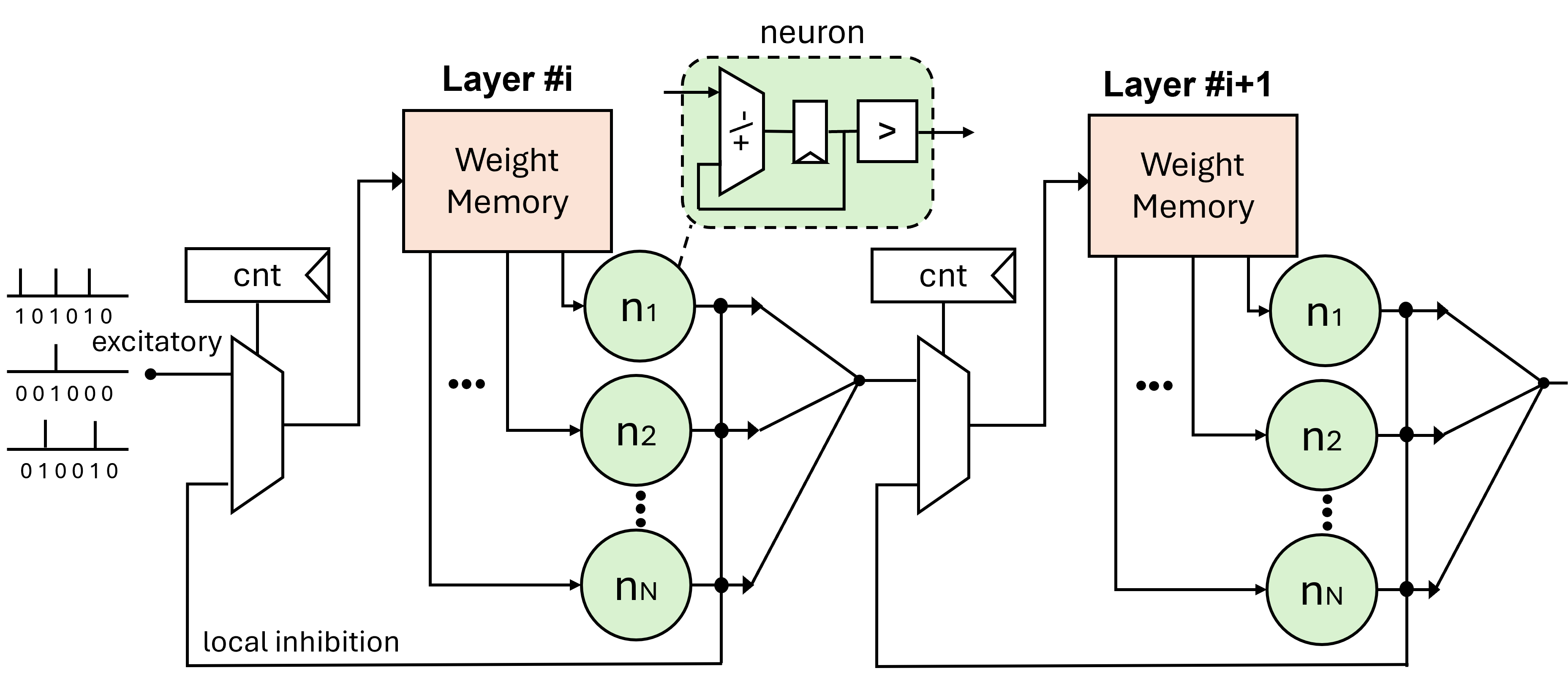}
\caption{Spatial dataflow SNN architecture template with dedicated neuron datapaths and weight memories for each layer.}
\label{f:snn-fpga}
\end{figure}

\subsection{Network Topology and Quantization}
High inference accuracy in SNNs often requires expressive network topologies. A common choice is a fully connected feed-forward network augmented with recurrent lateral inhibition, allowing active neurons to suppress their neighbors (Fig.~\ref{f:snn}). Although SNNs are typically trained using full-precision floating-point arithmetic, inference can be performed using finite-precision integer or fixed-point representations, avoiding the substantially higher hardware cost of floating-point operations~\cite{flowq}.

Quantization enables this transition by converting network variables into finite-precision representations, including synaptic weights, membrane potentials, firing thresholds, and leak calculations. For hardware implementation, quantization ultimately represents each synaptic weight as a finite-width binary value, reducing storage requirements while simplifying the arithmetic datapath. Further reductions in weight-memory requirements can be achieved through techniques such as pruning~\cite{pruning} and sparse weight representation~\cite{fpspike}, which reduce the number of weight values that must be stored.

\subsection{Synchronous Spatial-Dataflow FPGA  SNN Architecture}
To efficiently map SNN models onto an FPGA~\cite{s2n2, fpga-snn, modnef}, we employ a synchronous spatial-dataflow architecture like~\cite{spiker_plus}. As illustrated in Fig.~\ref{f:snn-fpga}, the complete network topology is instantiated in the FPGA, with each layer assigned a dedicated computational datapath and weight memory. The architecture follows a clock-driven execution model in which each layer processes its input spikes sequentially within a time step, followed by a state-evaluation phase.

By construction, exactly one active spike is handled per clock cycle, and an early-exit is allowed as soon as no active spikes remain across the inputs. For each selected active spike, the corresponding row of synaptic weights is read from the memory and broadcast to all neurons in the target layer, enabling their membrane potentials to be updated in parallel within the same cycle. Excitatory spikes from the preceding layer are processed first, followed by inhibitory spikes from within the same layer using the identical datapath. Because this sequential handling guarantees that no two input spikes arrive at a neuron simultaneously, the physical neuron architecture is drastically simplified. Effectively, each neuron is reduced to a basic adder or subtractor, performing a single arithmetic operation per active cycle to either add a synaptic weight for excitation or subtract one for inhibition.

Once all active input spikes for the current time step have been processed, the layer enters the state-evaluation phase. Membrane potentials are compared against $V_{\mathrm{th}}$ to generate output spikes. Firing neurons undergo a hard reset, followed by membrane leakage implemented using the aforementioned binary right shifts.

Execution proceeds to the next time step only after all layers have completed their state-evaluation phases.

\section{Hybrid Weight-Storage Architecture}

To reduce the substantial on-chip memory footprint of SNN accelerators, the proposed architecture replaces the conventional per-layer weight memory with a smaller memory that stores only a subset of the weight bits. The remaining bits are generated dynamically using combinational logic, reducing BRAM usage while preserving the original weight representation.

\begin{figure}[t]
\centering
\includegraphics[width=0.95\columnwidth]{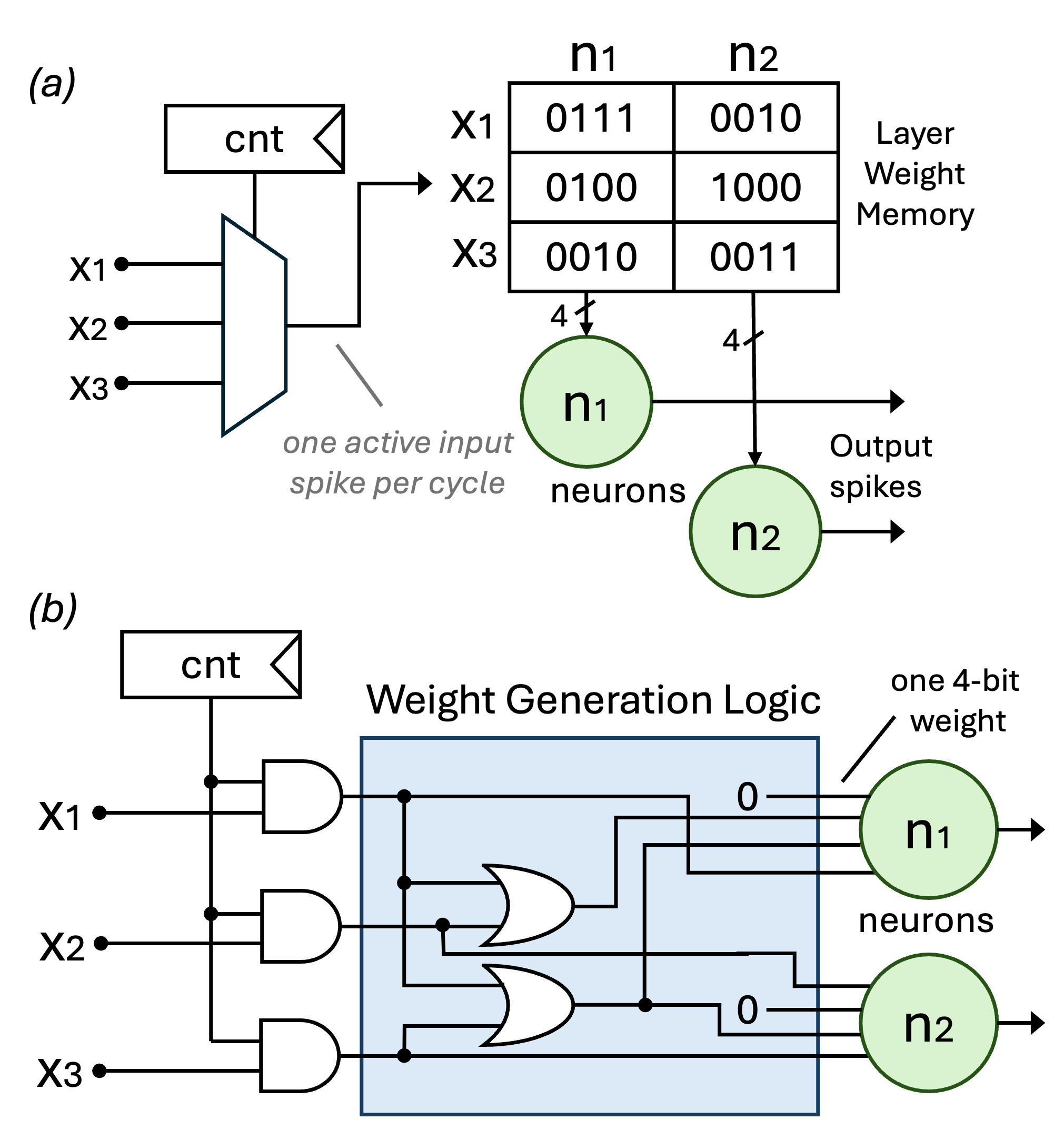}
\caption{Weight access for a layer with three input spikes and two neurons: (a) conventional BRAM-based weight fetching, and (b) combinational weight generation driven by at most one active input spike.}
\label{f:wgl}
\end{figure}

\subsection{Combinational weight-generation logic}

Consider a simplified layer with three input spikes ($x_1$, $x_2$, $x_3$) and two neurons ($n_1$, $n_2$). In the conventional memory-based design shown in Fig.~\ref{f:wgl}(a), the layer uses a weight memory with three rows. During the processing phase, when $x_1$ is active, the first row is read and its weights are broadcast to $n_1$ and $n_2$; similarly, the second and the third row is read when $x_2$ and $x_3$ is active, respectively.

For neuron $n_1$, activation of $x_1$ selects the 4-bit weight 0111, whereas activation of $x_2$ selects 0100 and $x_3$ selects 0010. Examining the individual bits across the three possible inputs reveals direct logical relationships:
\begin{itemize}
\item Bit 3 (MSB) remains `0' for all inputs and can therefore be hardwired to ground.
\item Bit 2 is `1' only when $x_1$ or $x_2$ are active.
\item Bit 1 is `1' for active inputs $x_1$ or $x_3$.
\item Bit 0 (LSB) is `1' only when $x_1$ is active.
\end{itemize}

As illustrated in Fig.~\ref{f:wgl}(b), these bit-level relationships translate directly into combinational logic for generating the weights of both neurons. The AND-mask gates and ring counter enforce the selection of a single input spike per clock cycle, so the logic only needs to generate the weights corresponding to the selected input. This produces the required weights for all neurons without storing them in memory.

However, generating all weight bits in logic becomes impractical for realistic layers with hundreds or thousands of inputs, as the resulting LUT network can incur substantial resource usage, routing congestion, and combinational delay. This scaling limitation motivates a hybrid implementation that combines LUT-based generation with conventional BRAM storage.

\subsection{Hybrid Weight-Storage Architecture}

\begin{figure}[htb]
\centering
\includegraphics[width=0.95\columnwidth]{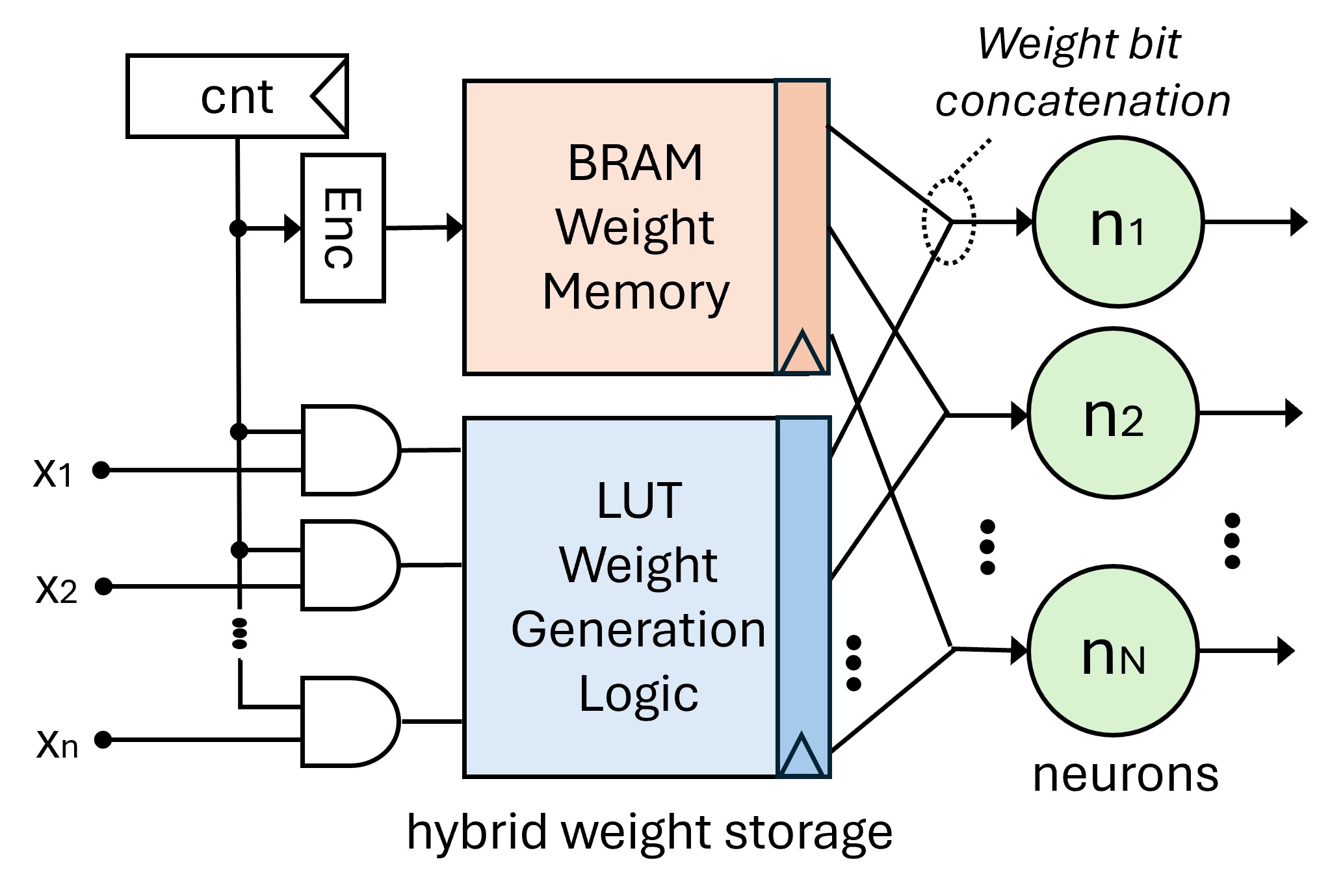}
\caption{Hybrid weight storage (BRAM) and online generation (LUTs) of the weights of an SNN layer.}
\label{f:hybrid}
\end{figure}

The resulting hybrid datapath combines the two sources of weight information during inference, as illustrated in Fig.~\ref{f:hybrid}. When an active spike is selected, its position is encoded and provided in parallel to the weight-generation logic and BRAM. The encoded position identifies the corresponding BRAM row, which provides the stored bits for all neurons in the target layer, while the weight-generation logic simultaneously produces the remaining bits. The two bit subsets are then concatenated at the neuron interface to reconstruct the original weight representation.

The same mechanism naturally supports recurrent intra-layer connections used for lateral inhibition, which require negative weights. To avoid storing or generating redundant sign bits, both BRAM and LUT-based logic provide only the unsigned weight magnitude. The sign is instead determined by the connection type at the datapath: excitatory spikes trigger an addition to the membrane potential, while inhibitory spikes trigger a subtraction. This decouples weight magnitude from its sign while supporting both forward and recurrent connections with the same weight-storage mechanism.

\subsection{Heuristic Allocation of Weight Bits Across LUTs/BRAMs}

To effectively partition the neural weights between memory and logic, we employ a practical allocation strategy. Driven by a designer specified target for memory reduction, a heuristic evaluates and distributes the individual weight bits. It maps the bits with the lowest logic synthesis cost to LUTs, while the rest bits are stored in BRAM.

A layer with $N$ neurons, $L$ possible input spikes, and $W$-bit weights contains $N \times L \times W$ bits in total. During each cycle, one of the $L$ input spikes is selected, and each of the $N$ neurons receives its corresponding, independently valued $W$-bit weight for that input. Thus, each cycle requires $N \times W$ weight bits, which can be independently sourced from BRAM, LUT-based logic, or a combination of both.

The logic-generation cost of each of the $N \times W$ weight bits is determined by the number of input-dependent OR terms required to generate it. With only one input spike active per cycle, a generated bit is the logical OR of the selected input signals for which that bit is `1' in the corresponding weight. Thus, the number of OR terms directly reflects the number of input dependencies and provides a measure of the logic cost.

\begin{figure}[h]
\centering
\includegraphics[width=0.9\columnwidth]{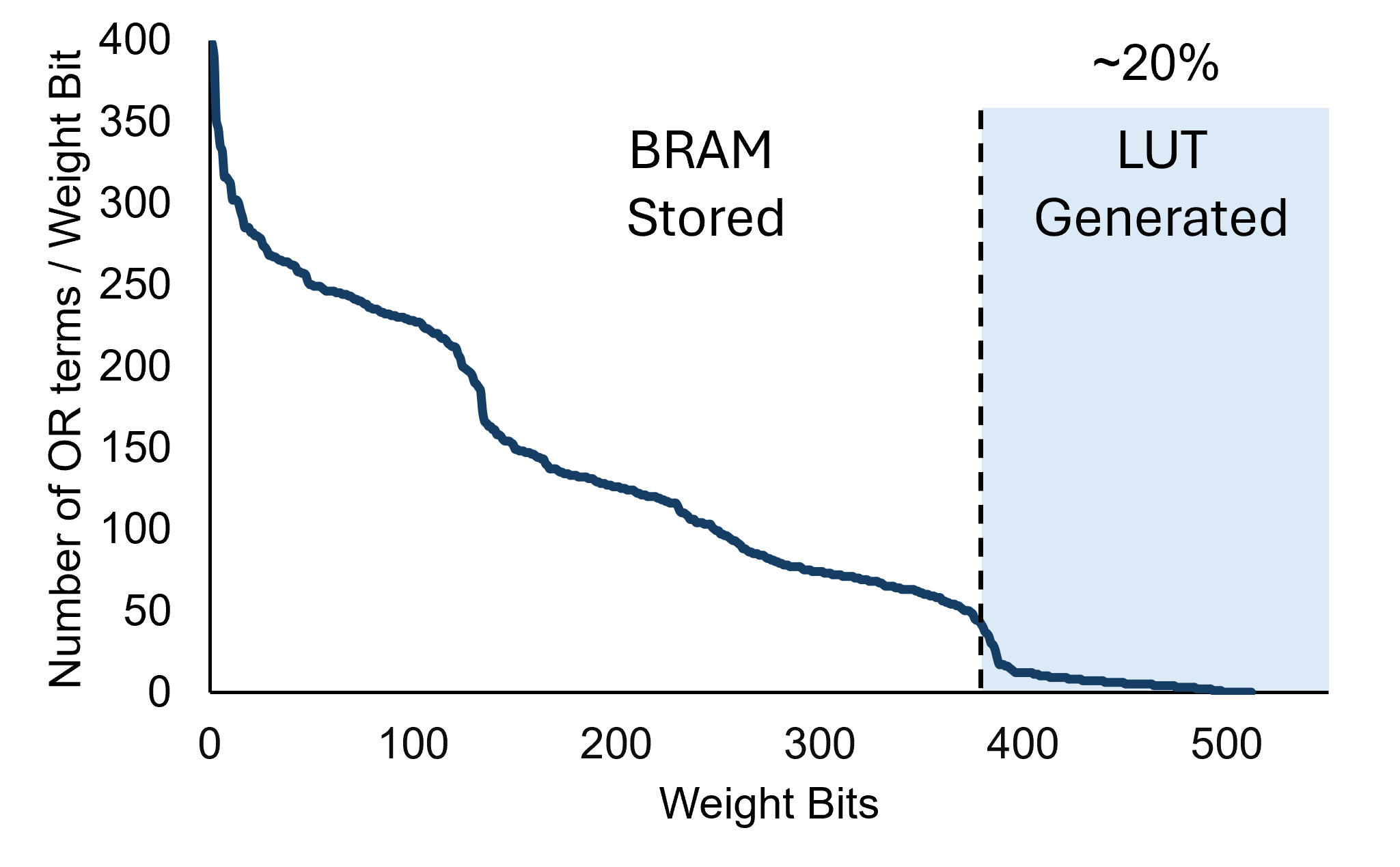}
\caption{Logic-generation cost of the $N\times W=128\times4=512$ weight bits for an MNIST layer, measured by the number of required OR terms. Bits are sorted by decreasing cost where the lowest-cost bits are mapped to LUTs, while the remainder is stored in BRAM.}
\label{f:bit-cost}
\end{figure}

Fig.~\ref{f:bit-cost} illustrates these costs for an MNIST layer ($N=128$, $W=4$). Each point represents one of the $N \times W$ weight bits, with the $y$-axis showing the number of OR terms required for its generation. The bits are sorted by this cost, from the lowest-cost bits on the right to the highest-cost bits on the left. For a target LUT allocation of 20\%, for example, the heuristic selects the 20\% of bits at the low-cost end of this distribution for LUT-based generation, while the remaining bits are stored in BRAM. This selected subset then forms the input to the weight-generation logic. Although logic sharing across bit positions could potentially reduce the cost of higher-dependency bits, our experiments show that selecting bits with the fewest OR terms provides an effective mapping. Assigning additional bits to LUTs can increase routing congestion and compromise timing closure. The current implementation considers only single-cycle, unpipelined weight-generation logic.

After selecting the bits for LUT-based generation, their Boolean expressions are optimized using Paar's algorithm~\cite{paar}. The algorithm identifies common input terms across bit expressions and recursively groups them into larger subexpressions, reducing the number of required logic gates. This optimization is applied before standard logic synthesis to obtain an efficient hardware implementation.

\section{Experimental evaluation}

\subsection{Evaluation Setup}
To evaluate the overall resource utilization and efficiency of the proposed architecture, we benchmark our implementation across three datasets: MNIST~\cite{mnist}, AudioMNIST~\cite{audiomnist}, and SHD~\cite{shd}. All network models are trained using the SNNTorch library~\cite{snntorch}, with full training details, hyperparameters, and network topology specifics are provided openly in~\cite{git-project}.

\begin{table*}
\centering 
\caption{Results for MNIST SNN models using the proposed hybrid BRAM/LUT weight storage compared with state-of-the-art FPGA accelerators.} 
\setlength{\tabcolsep}{5pt}
\renewcommand{\arraystretch}{1.2}
\begin{tabular}{|c||c|c|c|c|c||c|c|c|}\hline
Dataset & \multicolumn{5}{c||}{MNIST} & \multicolumn{3}{c|}{MNIST Large}\\ \hline
\multirow{4}{*}{Implementation} & Proposed & \multirow{4}{*}{\shortstack{Spiker+\\[-1pt]\cite{spiker_plus}}} & \multirow{4}{*}{\shortstack{ModNEF\\[-1pt]\cite{modnef}}} & \multirow{4}{*}{\shortstack{TripleBrain\\[-1pt]\cite{triplebrain}}} & \multirow{4}{*}{\shortstack{MorphBungee\\Tiny\\[-1pt]\cite{morph_tiny}}} & Proposed & \multirow{4}{*}{\shortstack{ModNEF\\[-1pt]\cite{modnef}}} & \multirow{4}{*}{\shortstack{\cite{edge_processor}}}\\ 
& L1: 23\% & & & & & L1: 23\% & & \\
& L2: 100\% & & & & & L2: 22\% & & \\  
& & & & & & L3: 100\% & & \\ \hline
Zynq FPGA & 7020 & 7020 & 7010 & 7045 & 7010 & 7020 & 7010 & 7045\\ \hline
SNN topology & 784-128-10 & 784-128-10 & 784-128-10 & 784-256 & 256-256-10 & 784-392-196-10 &	784-392-196-10 & 784-512-100 \\ \hline
Acc (\%) & 95.60 & 93.85 & 95.76 & 95.10 & 97.55 & 94.50 & 96.96 & 97.21 \\ \hline
Network Type & FC-FF-R & FC-FF & FC-FF & FC-FF & FC-FF & FC-FF-R & FC-FF & FC-FF\\ \hline
Potential bits & 11 & 6 & N.R.$^a$ & 24 & 16 & 12 & 16 & L1(13), L2(21)\\ \hline
Weight Bits & 4 & 4 & 10 & 16 & 12 & 4 & 8 & L1(8), L2(16)\\ \hline
LUTs & 7794 & 4314 & 5947 & 10052 &	8755 & 31263 & 20321 & 17578\\ \hline
FFs & 4466 & 3298 & 3060 & 8505 & 2970 & 14521 & 28284 & 3877\\ \hline
DSPs & - & - & - & 32 & - & - & 3 & -\\ \hline
BRAM36 & 11 & 18 & 42 & 131 & 25 & 84 & 113 & 137\\ \hline
BRAM18 & - & - & 1 & - & 1 & - & 1 & 1 \\ \hline
Power (mW) & 134 & 180 & 141 & 938 & 139 & 193 & 336 & 398\\ \hline
Energy (mW/MHz)	& 1.34 & 1.80 & 1.12 & 3.75 & 1.39 & 1.93 & 2.68 & 3.98\\ \hline
Frequency (MHz)	& 100 &	100 & 125 &	250 & 100 & 100 & 125 & 100\\ \hline
Latency/image (ms) & 0.259 & 0.780 & 0.115 & 0.370 & 0.253 & 0.467 & 0.904 & 0.467\\ \hline
\end{tabular}
\\$^a$N.R.: Not Reported 
\label{tab:Table1} 
\end{table*}

We compare the proposed design against a selection of highly resource efficient, state of the art FPGA SNN accelerators corresponding to each specific model case. All hardware utilization and inference accuracy metrics for these baseline architectures are reported directly as published in their respective papers:
\begin{itemize}
\item
Small MNIST Models: Our implementation is benchmarked against Spiker+~\cite{spiker_plus} and ModNef~\cite{modnef}, which employ similar network architectures on the same FPGA platform, as well as TripleBrain~\cite{triplebrain} and MorphBungee-Tiny~\cite{morph_tiny}.
\item
Large MNIST Models: We compare against ModNef~\cite{modnef} and the accelerator proposed by Chen et al.~\cite{edge_processor}.
\item 
Temporal Workloads (AudioMNIST~\cite{audiomnist} and SHD~\cite{shd}): To further assess the effectiveness of the architecture across diverse, time varying datasets, we extend our evaluation and compare the results directly against Spiker+~\cite{spiker_plus}.
\end{itemize}

Across these comparisons, we evaluate SNN model accuracy and relevant hardware metrics, including resource utilization (\#LUTs, \#FFs, BRAMs), power, energy efficiency (mW/MHz), achieved clock frequency, and latency per input. The proposed FPGA SNNs are designed in SystemVerilog RTL, open sourced in~\cite{git-project} under a permissible license, and synthesized for AMD Zynq-7020 SoC. Synthesis, implementation, and power evaluation are conducted using Xilinx Vivado. Reported power is the average power derived from post-implementation simulation using a representative, class-balanced subset of each SNN model's test set.

\subsection{Comparisons}

Table~\ref{tab:Table1} summarizes the results for the small and large MNIST models, with the proposed approach reported in the first column and state-of-the-art FPGA accelerators in the remaining columns. The proposed design uses the L1:23\%, L2:100\% configuration, meaning that 23\% of the weight bits in the first layer and all weight bits in the second layer are generated online using LUTs. This configuration represents a practical design point that provides substantial BRAM savings while maintaining balanced FPGA resource utilization.

For each implementation, the table reports the complete SNN topology, including the number of input connections, the sizes of the primary and secondary layers, and the data widths of the membrane potentials and weights. Spiker+ and ModNEF use the same network topology as the proposed design, while the remaining implementations employ topologies of comparable complexity. Despite the different hardware organizations, the proposed design maintains competitive inference accuracy across all designs.

The main benefit of the proposed approach is its substantial reduction in BRAM usage. In the selected configuration, it reduces the number of BRAMs by 38\% compared with Spiker+ and by up to 91\% compared with TripleBrain. This reduction comes at the cost of 17\% higher LUT/FF utilization on average than the state-of-the-art FPGA accelerators, representing a favorable tradeoff when on-chip BRAM is the primary resource constraint.

The reduced memory footprint also translates into lower power consumption, with the proposed design achieving the lowest power across all evaluated implementations; absolute power values are directly comparable for implementations targeting the same or smaller FPGA devices like in the case of Zynq 7020 and 7010, respectively. Energy efficiency (mW/MHz) follows a similar trend, although ModNEF performs slightly better. In terms of latency/input, ModNEF achieves the lowest average latency per input, while the proposed design closely follows, matching the performance of MorphBungee-Tiny.

The benefits of the proposed architecture become more pronounced as SNN size increases, where on-chip BRAM becomes a growing constraint. Table~\ref{tab:Table1} reports results for larger MNIST models too. The proposed design and ModNEF use the same topology, consisting of three equally sized layers, while~\cite{edge_processor} uses two larger layers.
The proposed design exhibits a slight accuracy degradation relative to the other models, primarily due to the training process rather than the hardware architecture. 
The proposed design reduces BRAM utilization by 26\% compared with ModNEF and by 39\% compared with~\cite{edge_processor}, while also using fewer combined LUTs and FFs than ModNEF.
It also achieves the lowest power consumption and best energy efficiency, while matching or improving latency relative to the compared designs.

Table~\ref{tab:Table3} compares the proposed architecture with Spiker+ on the AudioMNIST and SHD datasets. The comparison is particularly relevant as both designs target the same FPGA platform and operate at the same clock frequency of 100 MHz, using similar network topologies. The proposed design achieves 2.78\% higher accuracy on AudioMNIST, while trailing Spiker+ by only 0.8\% on SHD. Recurrent connections are used in all configurations except the AudioMNIST Spiker+ model.

The proposed architecture achieves the lowest BRAM utilization on both datasets, with reductions of 43\% for AudioMNIST and 39\% for SHD. These savings require 20\% and 17\% more LUTs/FFs, respectively. The reduced memory footprint also translates into the lowest power consumption for both datasets.
For latency, Spiker+ performs slightly better on AudioMNIST, while the two designs achieve nearly identical latency on SHD.

\begin{table}[!t] \centering 
\caption{Experimental results of the proposed SNN against Spiker+ on the AUDIO MNIST and SHD dataset.} 
\setlength{\tabcolsep}{2.8pt}
\renewcommand{\arraystretch}{1.2}
\begin{tabular}{|c||c|c||c|c|}\hline
Dataset & \multicolumn{2}{c||}{AudioMNIST} & \multicolumn{2}{c|}{SHD} \\ \hline
\multirow{3}{*}{Implementation} & Proposed & \multirow{3}{*}{\shortstack{Spiker+\\[-1pt]\cite{spiker_plus}}} & Proposed & \multirow{3}{*}{\shortstack{Spiker+\\[-1pt]\cite{spiker_plus}}} \\ 
& L1: 28\% & & L1: 20\% & \\ 
 & L2: 100\% & & L2: 100\% & \\ \hline
Zynq FPGA & 7020 & 7020 & 7020 & 7020 \\ \hline
SNN topology & 40-150-10 & 40-150-10 & 700-200-20 &	700-200-20 \\ \hline
Acc (\%) & 92.60 & 89.82 & 72.20 & 72.99 \\ \hline
Network Type & FC-FF-R & FC-FF & FC-FF-R & FC-FF-R \\ \hline
Potential bits & 12 & 8 & 14 & 8 \\ \hline
Weight Bits & 6 & 5 & 7 & FF(6)/R(5) \\ \hline
LUTs & 8603 & \multirow{2}{*}{10124} & 14725 & \multirow{2}{*}{18268} \\ \hhline{--~-~}
FFs & 3598 & & 6675 & \\ \hline
DSPs & - & - & - & - \\ \hline
BRAM36 & 9 & 16 & 31 & 51 \\ \hline
BRAM18 & - & - & - & - \\ \hline
Power (mW) & 146 & 290 & 151 & 430 \\ \hline
Energy (mW/MHz) & 1.46 & 2.90 & 1.51 & 4.30 \\ \hline
Frequency (MHz)	& 100 & 100 & 100 & 100 \\ \hline
Latency/recording (ms) & 0.329 & 0.080 & 0.574 &	0.540 \\ \hline
\end{tabular}
\label{tab:Table3} 
\end{table}

 Overall, the evaluations confirm that the proposed hybrid weight storage approach successfully trades memory capacity for programmable logic. This architectural shift enables highly memory efficient SNN acceleration while sustaining competitive accuracy, power consumption, and latency across varied network sizes and datasets.

\subsection{Limits of Hybrid Weight Storage}
\label{sec:max_reductions}
To assess how far this tradeoff can be pushed, Table~\ref{tab:Table4} reports the proposed configurations achieving the maximum BRAM reduction for each SNN model while maintaining routability and the target clock frequency. Beyond these points, further BRAM reduction results in either unroutable designs or failure to achieve timing closure at 100 MHz.

The achieved BRAM reduction is substantial, cutting memory usage by approximately half compared with the balanced configurations reported in Tables~\ref{tab:Table1}--\ref{tab:Table3}, which already provide significant savings over the state-of-the-art. However, this aggressive reduction inevitably increases the size of the weight-generation logic, partially offsetting the power benefits of reduced memory usage. Although the designs in Table~\ref{tab:Table4} still consume less power than the state of the art in most cases, their power consumption is marginally higher, by less than 2.5\%, than that of our balanced configurations. Thus, the memory-to-logic tradeoff must be carefully managed to avoid excessive combinational growth.

More broadly, Table~\ref{tab:Table4} demonstrates that the proposed architecture can adapt to the available FPGA resources by shifting the weight-storage requirement between BRAM and LUTs. This flexibility is particularly valuable for resource-constrained devices, where BRAM capacity can limit the size of deployable SNNs even when programmable logic remains available. The proposed approach therefore provides a practical means of extending the range of SNN models that can be accommodated by a given FPGA.

\begin{table}[!t] 
\centering 
\caption{The proposed designs on a Zynq 7020 at 100MHz with the maximum BRAM savings for all examined SNN models.}
\setlength{\tabcolsep}{4pt}
\renewcommand{\arraystretch}{1.2}
\begin{tabular}{|c|c|c|c|c|}\hline
Dataset & MNIST & MNIST Large & AudioMNIST & SHD\\ \hline
\multirow{3}{*}{Implementation} & L1: 61\% & L1: 31\% & L1: 80\% & L1: 36 \% \\ 
&  L2: 100\% & L2: 31\% & L2: 100\% & L2: 100\% \\ 
& & L3: 100\% & & \\ \hline
LUTs & 10561 & 32450 & 12433 & 17423 \\ \hline
FFs & 4663 & 14719 & 4067 & 6886 \\ \hline
BRAM36 & 5 & 75 & 2 & 25 \\ \hline
BRAM18 & 1 & - & 1 & - \\ \hline
Power (mW) & 137 & 193 & 146 & 153 \\ \hline
\end{tabular}
\label{tab:Table4} 
\end{table}
\section{Conclusions and Future Work}
In this work, we introduced a novel architectural paradigm for digital FPGA SNNs that dynamically generates a targeted portion of synaptic weights directly within programmable logic. Supported by a heuristic partitioning strategy that evaluates logic synthesis costs, this method efficiently distributes individual weight bits across available hardware resources (i.e., BRAMs and LUTs). The resulting implementations achieve significant reductions in memory footprint and overall power consumption without compromising inference accuracy or processing throughput.

Moving forward, this hybrid framework offers clear pathways for further enhancement. Future research will explore hardware and algorithm co-design, specifically targeting custom weight encodings and specialized training constraints. For instance, snapping weights to values with naturally sparse bit patterns, like powers of two, can actively minimize the required density of logical ones. Additionally, the allocation algorithm can be expanded by incorporating physical synthesis and routing metrics directly into the selection loop, pushing the optimization of the logic to memory distribution even further.

\bibliographystyle{IEEEtran}
\bibliography{refs}

@INPROCEEDINGS{layerspec,
  author={Lebdeh, Muath Abu and Yildirim, Kasim Sinan and Brunelli, Davide},
  booktitle={Intern. Conf. on Neuromorphic Systems (ICONS)}, 
  title={Per Layer Specialization for Memory-Efficient Spiking Neural Networks}, 
  year={2024},
  pages={169-176},}

@inproceedings{paar,
  author    = {Paar, Christof},
  title     = {Optimized arithmetic for Reed-Solomon encoders},
  booktitle = {Proc. of IEEE International Symposium on Information Theory},
  pages     = {250},
  year      = {1997},
}

@inproceedings{neuralut,
  title={Neuralut: Hiding neural network density in boolean synthesizable functions},
  author={Andronic, Marta and Constantinides, George A},
  booktitle={Intern, Conf. on Field-Programmable Logic and Applications (FPL)},
  pages={140--148},
  year={2024},
}

@inproceedings{pedram,
  title={Energy-efficient, low-latency realization of neural networks through boolean logic minimization},
  author={Nazemi, Mahdi and Pasandi, Ghasem and Pedram, Massoud},
  booktitle={Proc. of the Asia and South Pacific design automation conference (ASPDAC)},
  pages={274--279},
  year={2019}
}

@INPROCEEDINGS{loihi2,
author={Shrestha, Sumit Bam and Timcheck, Jonathan and Frady, Paxon and Campos-Macias, Leobardo and Davies, Mike},
booktitle={IEEE International Conference on Acoustics, Speech and Signal Processing (ICASSP)}, 
title={Efficient Video and Audio Processing with Loihi 2}, 
year={2024},
pages={13481-13485},
}

@article{spinnaker2,
  title={The SpiNNaker2 chip: a many-core platform for flexible and scalable brain-inspired computing},
  author={Scholze, Stefan and Partzsch, Johannes and H{\"o}ppner, Sebastian and Kelber, Florian and Dixius, Andreas and Stolba, Marco and Arfa, Sirine and Berthel, Marc and Ellguth, Georg and Garside, Jim and others},
  journal={IEEE Open Journal of Circuits and Systems},
  year={2026},
}

@article{spiker_plus,  
 title={Spiker+: a framework for the generation of efficient Spiking Neural Networks FPGA accelerators for inference at the edge},  
 author={Carpegna, Alessio and Savino, Alessandro and Di Carlo, Stefano},  
 journal={IEEE Trans. on Emerging Topics in Computing},  
 volume={13},  
 number={3},  
 pages={784--798},  
 year={2024},  
 publisher={IEEE}}

@article{modnef,
  title={Modnef: An open source modular neuromorphic emulator for {FPGA} for low-power in-edge artificial intelligence},
  author={Saulquin, Aur{\'e}lie and others},
  journal={ACM TACO},
  volume={22},
  number={2},
  pages={1--24},
  year={2025},
}

@ARTICLE{firefly,
  author={Li, Jindong and others},
  journal={IEEE Trans. on CAD}, 
  title={{FireFly} v2: Advancing Hardware Support for High-Performance Spiking Neural Network With a Spatiotemporal {FPGA} Accelerator}, 
  year={2024},
  volume={43},
  number={9},
  pages={2647-2660},
}

@article{fpspike,
  title={Fpspike: A fully parallel and reconfigurable architecture for accelerating spiking neural networks with structured sparsity},
  author={Li, Mingyang and Kan, Yirong and Wu, Man and Zhang, Renyuan and Nakashima, Yasuhiko},
  journal={IEEE Trans. on Circuits and Systems I: Regular Papers},
  year={2026},
}

@inproceedings{s2n2,
  title={S2n2: A {FPGA} accelerator for streaming spiking neural networks},
  author={Khodamoradi, Alireza and Denolf, Kristof and Kastner, Ryan},
  booktitle={ACM Intern. Symp. on Field-Programmable Gate Arrays},
  pages={194--205},
  year={2021}
}

@article{mnist,
  title={Gradient-based learning applied to document recognition},
  author={LeCun, Yann and others},
  journal={Proc. of the IEEE},
  volume={86},
  number={11},
  pages={2278--2324},
  year={1998},
  publisher={Ieee}
}

@article{shd,
  title={The heidelberg spiking data sets for the systematic evaluation of spiking neural networks},
  author={Cramer, Benjamin and others},
  journal={IEEE Trans. on Neural Networks and Learning Systems},
  volume={33},
  number={7},
  pages={2744--2757},
  year={2020},
  publisher={IEEE}
}

@article{audiomnist,
  title={Audiomnist: Exploring explainable artificial intelligence for audio analysis on a simple benchmark},
  author={Becker, S{\"o}ren and others},
  journal={Journal of the Franklin Institute},
  volume={361},
  number={1},
  pages={418--428},
  year={2024},
  publisher={Elsevier}
}

@INPROCEEDINGS{flowq,
  author={Asim, Faaiz and Aung, Sanhtet and Lee, Jongeun},
  booktitle={ASP-DAC}, 
  title={{FlowQ:} Fixed-point Low-precision Post-Training Quantization Framework for Efficient and Accurate SNN Inference}, 
  year={2026},
  pages={119-125},
}

@article{fpga-snn,
author={Mehrzad Karamimanesh and others},
title = {Spiking neural networks on {FPGA}: A survey of methodologies and recent advancements},
journal = {Neural Networks},
volume = {186},
pages = {107256},
year = {2025},
}

@article{pruning,
  title={Pruning and quantization for deep neural network acceleration: A survey},
  author={Liang, Tailin and others},
  journal={Neurocomputing},
  volume={461},
  pages={370--403},
  year={2021},
  publisher={Elsevier}
}

@article{snntorch,
        title   =  {Training spiking neural networks using lessons from deep learning},
        author  =  {Eshraghian, Jason K and others},
        journal = {Proc. of the IEEE},
        volume  = {111},
        number  = {9},
        pages   = {1016--1054},
        year    = {2023}
}

@inproceedings{vq4snn,
    author = {Dimitrios Sekertzis and Giorgos Dimitrakopoulos},
    title = {{VQ4SNN}: Vector Quantization for Memory-Efficient {FPGA} Spiking Neural Networks},
    booktitle = {IEEE International Conference on Electronics Circuits and Systems (ICECS)},
    year = 2026
}

@misc{git-project,
  author       = {{IC-Lab-DUTH}},
  title        = {{FPGA SNN with hybrid BRAM-LUT weight storage}},
  year         = {2026},
  howpublished = {www.github.com/ic-lab-duth/hybrid-weight-storage-snn.git},
}

@article{snn-vs-ann,
  title={Are SNNs really more energy-efficient than ANNs? An in-depth hardware-aware study},
  author={Dampfhoffer, Manon and Mesquida, Thomas and Valentian, Alexandre and Anghel, Lorena},
  journal={IEEE Trans. on Emerging Topics in Computational Intelligence},
  volume={7},
  number={3},
  pages={731--741},
  year={2022},
}

@article{survey,
  title={A survey of neuromorphic computing and neural networks in hardware},
  author={Schuman, Catherine D and Potok, Thomas E and Patton, Robert M and Birdwell, J Douglas and Dean, Mark E and Rose, Garrett S and Plank, James S},
  journal={arXiv preprint arXiv:1705.06963},
  year={2017}
}

@INPROCEEDINGS{spiker,
  author={Carpegna, Alessio and Savino, Alessandro and Di Carlo, Stefano},
  booktitle={2022 IEEE Computer Society Annual Symposium on VLSI (ISVLSI)}, 
  title={Spiker: an FPGA-optimized Hardware accelerator for Spiking Neural Networks}, 
  year={2022},
  volume={},
  number={},
  pages={14-19},
  doi={10.1109/ISVLSI54635.2022.00016}}

@ARTICLE{triplebrain,
  author={Wang, Haibing and He, Zhen and Wang, Tengxiao and He, Junxian and Zhou, Xichuan and Wang, Ying and Liu, Liyuan and Wu, Nanjian and Tian, Min and Shi, Cong},
  journal={IEEE Trans. on Biomedical Circuits and Systems}, 
  title={TripleBrain: A Compact Neuromorphic Hardware Core With Fast On-Chip Self-Organizing and Reinforcement Spike-Timing Dependent Plasticity}, 
  year={2022},
  volume={16},
  number={4},
  pages={636-650},
  doi={10.1109/TBCAS.2022.3189240}}

@ARTICLE{morph_tiny,
  author={Zhong, Zhengqing and Yang, Jialin and Wang, Haibing and Jiang, Ying and Lin, Yingcheng and Gan, Ping and Liu, Liyuan and Liang, Liang and Shi, Cong},
  journal={IEEE Trans. on Circuits and Systems II: Express Briefs}, 
  title={MorphBungee-Tiny: An Ultra-Low-Cost FPGA Implementation of Neuromorphic Architecture With High-Accuracy On-Chip E-Prop Learning}, 
  year={2025},
  volume={72},
  number={12},
  pages={2012-2016},
  doi={10.1109/TCSII.2025.3623997}}

@ARTICLE{edge_processor,
  author={Chen, Guanyu and Chen, Mingju and Wang, Tengxiao and Wang, Haibing and Fu, Xiang and Lin, Yingcheng and Liu, Liyuan and Shi, Cong},
  journal={IEEE Trans. on Circuits and Systems II: Express Briefs}, 
  title={An Edge Neuromorphic Processor With High-Accuracy On-Chip Aggregate-Label Learning}, 
  year={2025},
  volume={72},
  number={3},
  pages={509-513},
  doi={10.1109/TCSII.2025.3529670}}

@INPROCEEDINGS{memory_review,
  author={Le, Dexter and Arig, Baran and Isik, Murat and Dikmen, I. Can and Karadag, Teoman},
  booktitle={IEEE Intern. Conf. on Computing and Machine Intelligence (ICMI)}, 
  title={A Review of Memory Wall for Neuromorphic Computing}, 
  year={2025},
  pages={1-6},
}

@ARTICLE{fpga_survey,
  author={Farsa, Edris Zaman and Ahmadi, Arash and Keszocze, Oliver},
  journal={IEEE Transactions on Emerging Topics in Computational Intelligence}, 
  title={Reconfigurable Digital FPGA Implementations for Neuromorphic Computing: A Survey on Recent Advances and Future Directions}, 
  year={2025},
  volume={9},
  number={5},
  pages={3210-3232},
  doi={10.1109/TETCI.2025.3551934}}

\end{document}